\documentclass[preprint,preprintnumbers,amssymb,prd,nofootinbib,tightenlines]{revtex4}
\usepackage{graphicx}
\usepackage{dcolumn}
\usepackage{bm}

\newcommand{\be}{\begin{equation}}
\newcommand{\ee}{\end{equation}}
\newcommand{\ba}{\begin{eqnarray}}
\newcommand{\ea}{\end{eqnarray}}

\begin{document}

\preprint{SLAC-PUB-9162}
\preprint{UK/TP-2002-03} 

\title{Direct CP Violation in Untagged $B$-Meson Decays}

\author{S. Gardner}
\email{gardner@pa.uky.edu}

\affiliation{
Stanford Linear Accelerator Center, 
Stanford University, Stanford, CA 94309
}
\affiliation{
Department of Physics and Astronomy, University of Kentucky, 
Lexington, KY 40506-0055\footnote{Permanent Address.}
}


\vskip 1.0 cm

\begin{abstract}
Direct CP violation can exist in 
untagged, neutral $B$-meson decays to 
certain self-conjugate, hadronic final
states. It can occur if the resonances which appear 
therein permit the identification of 
distinct, CP-conjugate states --- in analogy to stereochemistry, 
we term such states ``CP-enantiomers.'' 
These states permit the construction of 
a CP-odd amplitude combination in the untagged decay rate, which 
is non-zero if direct CP violation is present. 
The decay $B\to \pi^+\pi^-\pi^0$, containing the distinct CP-conjugate
states $\rho^+ \pi^-$ and $\rho^- \pi^+$, provides one such 
example of a CP-enantiomeric pair. 
We illustrate the possibilities in various multi-particle final
states.
\end{abstract}


\maketitle

The measurement of a non-zero value of 
${\rm Re}(\epsilon^\prime/\epsilon)$ in $K\to \pi\pi$ decays
establishes the existence of 
direct CP violation in nature~\cite{epsprime}, 
and provides an important first check
of the mechanism of CP violation in the Standard Model (SM). 
Numerically, however, ${\rm Re}(\epsilon^\prime/\epsilon)$ 
is very small. 
In the SM, this results, in part, from the 
weakness of inter-generational mixing~\cite{bosch}; the associated 
CP-violating parameter $\delta_{KM}$ in the 
Cabibbo-Kobayashi-Maskawa (CKM)
matrix need not be small~\cite{ckmmatrix}. 
Indeed, 
the measurement of a large CP-asymmetry 
in $B^0 (\bar B^0) \to J/\psi K_s$ decay and related modes~\cite{psiks}, 
induced through 
the interference of $B^0-\bar B^0$ mixing and direct decay, 
suggests that $\delta_{KM}\sim {\cal O}(1)$~\cite{nir}. 
Nevertheless, the observation of direct CP violation in the 
$B$-meson system is needed to clarify the mechanism of CP violation, 
to confirm that the Kobayashi-Maskawa (KM) phase drives the 
CP-violating effects seen. 
In the SM, direct CP violation is anticipated to be much larger
in $B$-meson decays than in $K$-meson decays~\cite{bander}. 
The observation of direct CP violation in $B$-meson decays would
falsify models in which the CP-violating interactions
are ``essentially'' superweak~\cite{superweak,superweak2}. 
In this paper, we discuss how the presence of direct CP violation
can be elucidated in untagged $B$-meson decays --- the practical 
advantage of this strategy is the far larger statistical sample of
events available.

The rich resonance structure of the 
multiparticle ($n\ge 2$) final states accessible 
in heavy meson decays
provides the possibility of observing direct CP violation 
without tagging the flavor of the decaying, neutral meson. The familar
condition for the presence of direct CP violation, 
$|\bar A_{\bar f}/A_f|\ne 1$, can be met by a non-zero value of the
partial rate asymmetry, so that,  
seemingly, one would 
want to distinguish empirically a decay with amplitude $A_f$ from that of its
CP-conjugate mode with amplitude $\bar A_{\bar f}$. 
However, in neutral $B$, $D$-meson decays to 
self-conjugate final states~\cite{Carter,Bigi,Dunietz:1986vi}, 
direct CP violation in untagged decays may nevertheless occur. 
It can occur if we can separate the self-conjugate final state, via
the resonances which appear, 
into distinct, CP-conjugate states. 
This condition finds it analogue in stereochemistry: we refer
to molecules which are non-superimposable, mirror images of each
other as enantiomers~\cite{ch41}. Accordingly, we refer to 
non-superimposable, 
CP-conjugate states as {\it CP enantiomers}. 
In $B\to \pi^+\pi^-\pi^0$ decay, e.g., 
the intermediate
states $\rho^+\pi^-$ and $\rho^-\pi^+$ form CP enantiomers, as
they are distinct, CP-conjugate states. As a result, 
the untagged decay rate contains a CP-odd amplitude combination.
The empirical presence of this CP-odd interference term in the untagged
decay rate would be realized in the Dalitz plot as a population asymmetry, 
reflective of direct CP violation.

We shall use $B\to \pi^+\pi^-\pi^0$ decay as a paradigm of how
direct CP violation can occur in untagged $B$-meson decays. 
In what follows, we shall largely follow the notation and
conventions of Quinn and Silva~\cite{QS}. 
Consider the amplitudes for $B^0 (\bar B^0) \to \pi^+ \pi^- \pi^0$ decay:
\ba
A(B^0(p_B)\to \pi^+(p_+)\pi^-(p_-)\pi^0(p_0))&=&
f_+[u]\, a_{+-} + f_-[s]\, a_{-+} + f_0[t]\, a_{00}\,, 
\nonumber\\
\bar A(\bar B^0(p_B)\to \pi^+(p_+)\pi^-(p_-)\pi^0(p_0))&=&
f_+[u]\, \bar a_{+-} + f_-[s]\, \bar a_{-+} + 
f_0[t]\, \bar a_{00} \;,
\ea
where the two-body decay amplitudes are given by 
$a_{+-}=A(B^0\to\rho^+ \pi^-)$, $a_{-+}=A(B^0\to\rho^- \pi^+)$,
and $a_{00}=A(B^0\to\rho^0 \pi^0)$ and $f_i$ is the 
form factor describing $\rho^i \to \pi \pi$. 
We have used $s=(p_- + p_0)^2$, $t=(p_+ + p_-)^2$, and 
$u=(p_+ + p_0)^2$ \footnote{We have implicitly summed over the
$\rho^i$ polarization. Defining 
$\langle \pi^0 (p_0) \pi^- (p_-)| \rho^- (p_\rho, \epsilon) \rangle
\equiv -g_\rho \, {\epsilon} \cdot (p_0 - p_-)$ and 
$\langle \rho^i(\epsilon, p_\rho) \pi^j(p_\pi)
|{\cal H}_{\rm eff} |  B^0(p_B) \rangle
\equiv 2\epsilon^\ast\cdot p_{\pi} a_{ij}$,  where 
${\cal H}_{\rm eff}$ is the $|\Delta B|=1$ effective Hamiltonian,
we find 
$A(B^0(p_B) \to \pi^+(p_+)\pi^-(p_-)\pi^0(p_0))
=  a^{00} (s - u)F_0(t)  + a^{+-} (t-s) F_+(u)
+ a^{-+} (u-t) F_-(s)$, where the pions' masses are
given by $M_{\pi^\pm}=M_{\pi^0}=M_{\pi}$. The form factor
$F_i(x)$ can be described by
a Breit-Wigner form $g_\rho/(x - M_\rho^2 + i \Gamma_\rho M_\rho)$, or
a more sophisticated function, consistent with the theoretical constraints
of analyticity, time-reversal-invariance, 
and unitarity, see Ref.~\protect{\cite{svgulf}}
for all details. Note that, e.g., $f_+[u] \equiv (t-s)F_+(u)$. 
}. 
For clarity, note
that $\bar a_{+-}=\bar A(\bar B^0\to\rho^+ \pi^-)$ and
$\bar a_{-+}=\bar A(\bar B^0\to\rho^- \pi^+)$.
Since $\rho^+ \pi^-$
and $\rho^-\pi^+$ are distinct, CP-conjugate states, 
the amplitudes 
$a_g= a_{+-} + a_{-+}$ and $a_u= a_{+-} - a_{-+}$ have distinct
properties under CP.
That is, if we define
$\bar a_g = \bar a_{+-} + \bar a_{-+}$ and 
$\bar a_u= \bar a_{+-} - \bar a_{-+}$, we see, under an
appropriate choice of phase conventions, that the CP conjugate
of $a_g$ is $\bar a_g$, whereas the CP conjugate of $a_u$ is 
$-\bar a_u$. With $a_n=2a_{00}$ we have 
\ba
A_{3\pi} &\equiv& A(B^0\to \pi^+\pi^-\pi^0) = 
f_g[u,s]\, a_{g} + f_u[u,s]\, a_{u} + f_n[t]\, a_{n}\; \nonumber \\
\bar A_{3\pi} &\equiv& \bar A(\bar B^0\to \pi^+\pi^-\pi^0) = 
f_g[u,s]\, \bar a_{g} + f_u[u,s]\, \bar a_{u} + f_n[t]\, \bar a_{n}\;,
\ea
where 
$f_g[u,s]= (f_+[u] + f_-[s])/{2}$, $f_u[u,s]= (f_+[u] - f_-[s])/{2}$, 
and $f_n[t]= {f_0[t]}/{2}$. 
Neglecting the width difference of the $B$-meson mass eigenstates, as
$\Delta \Gamma \equiv \Gamma_H - \Gamma_L$ and $|\Delta\Gamma| \ll
\Gamma\equiv (\Gamma_H + \Gamma_L)/2$, the decay
rate into $\pi^+\pi^-\pi^0$ for a $B^0$ meson at time $t=0$ is given
by~\cite{Azimov:ra}
\begin{equation}
\Gamma(B^0(t) \to \pi^+\pi^-\pi^0) = |A_{3\pi}|^2 e^{-\Gamma t}
\left[\frac{1 + |\lambda_{3\pi}|^2}{2} +
\frac{1 - |\lambda_{3\pi}|^2}{2} \cos(\Delta m\,t)
-{\rm Im}\lambda_{3\pi}\sin(\Delta m\,t) \right] \;,
\end{equation}
whereas the analogous decay rate for a $\bar B^0$
meson at time $t=0$ is given by 
\begin{equation}
\Gamma(\bar B^0(t) \to \pi^+\pi^-\pi^0) = |A_{3\pi}|^2 e^{-\Gamma t}
\left[\frac{1 + |\lambda_{3\pi}|^2}{2} -
\frac{1 - |\lambda_{3\pi}|^2}{2} \cos(\Delta m\,t)
+{\rm Im}\lambda_{3\pi}\sin(\Delta m\,t) \right]\;.
\end{equation}
Note that 
$\lambda_{3\pi} \equiv q \bar A_{3\pi}/p A_{3\pi}$  
and $\Delta m \equiv M_H - M_L$. We neglect $\Delta \Gamma$, so that we set 
$|q/p|=1$. Untagged observables, 
for which the identity of the $B$ meson
at $t=0$ is unimportant, correspond to 
$\Gamma(B^0(t) \to \pi^+\pi^-\pi^0) + \Gamma(\bar B^0(t) \to \pi^+\pi^-\pi^0)
\propto 
|A_{3\pi}|^2 + |\bar A_{3\pi}|^2$. We have 
\ba
|A_{3\pi}|^2 + |\bar A_{3\pi}|^2 &=&
\sum_i (|a_i|^2 + |\bar a_i|^2) |f_i|^2  \;\nonumber \\
&+& 2\sum_{i<j} \left[{\rm Re}(f_i f_j^*)\,{\rm Re}(a_i a_j^* + 
\bar a_i \bar a_j^*)
-  {\rm Im}(f_i f_j^*)\,{\rm Im}(a_i a_j^* + \bar a_i \bar a_j^*)\right] \;,
\label{sumAbarA}
\ea
where $i,j \in g,u,n$, noting that $i,j$ labels are not repeated
in the sum labelled ``$i<j$''. 
The different products $f_i f_j^*$ 
are distinguishable through the Dalitz plot of this decay, so that 
the coefficients of these functions are empirically distinct~\cite{QS}.
For our purposes the crucial point is that
these observables, as first noted by Quinn and Silva~\cite{QS}, 
can be of CP-odd character. In particular, the presence of 
\be
\label{direct}
a_g a_u^* + \bar a_g \bar a_u^* \quad \hbox{and/or} \quad
a_n a_u^* + \bar a_n \bar a_u^* 
\ee
is reflective of direct CP violation. Physically these observables
correspond to a population asymmetry 
under the exchange of $u$ and $s$ (or of $p_+$ and $p_-$)
across the Dalitz plot. To make the geometric
sense of this construction clear, consider a Dalitz plot in 
$u$ versus $s$, that is, in the invariant masses of the $\pi^+\pi^0$ and 
$\pi^-\pi^0$ pairs, respectively --- such a plot is shown in Fig.~1
of Ref.~\cite{babarb3pi}. 
The presence of the CP-odd amplitude 
$a_g a_u^* + \bar a_g \bar a_u^*$, e.g., engenders a population asymmetry
about the $u=s$ ``mirror line;'' specifically, the number of charged 
$\rho$ events in the $u> s$ region differs from that in the $s<u$ region.
Note that the functional form of $f_+(u)$ and 
$f_+(s)$ restrict the product $f_g f_u$ to the $\rho^{\pm}$ bands
in the Dalitz plot. The asymmetry is largest in the regions where the 
$\rho^i$ bands overlap, though the restricted number of events
in the overlap region make it more efficacious to compare the entire 
population of the charged $\rho$ bands 
in the $u>s$ and $u<s$ regions~\cite{sgjt}. 
The second amplitude combination of Eq.~(\ref{direct}) is determined
by the population asymmetry across the $u=s$ line 
in the regions in which the
$\rho^\pm$ and $\rho^0$ bands overlap. 
A population asymmetry in $B, \bar B \to \pi^+ \pi^- \pi^0$ decay 
about the $u=s$ line is also a signature of direct CP violation. 
However, non-zero values of the amplitude combinations of Eq.~(\ref{direct})
do not guarantee its existence as cancellations, though 
likely incomplete, can occur. 
The direct CP-violating observables of Eq.~(\ref{direct}) 
can persist even if the strong phases of the $a_j$ amplitudes
were zero. 
To illustrate, we parametrize
$a_j=T_j \exp(-i\alpha) + P_j$
and $P_j/T_j = r_j \exp(i\delta_j)$, where $r_j>0$ and $\delta_j$ is
the strong phase of interest\footnote{We drop an overall factor
of $\exp(-i\beta)$ in $a_j$ as it is of no consequence to our discussion.}. 
Thus
\be
\label{breakdown}
a_g a_u^* + \bar a_g \bar a_u^* 
= -2  
T_g T_{u}^{\ast}\, \sin\alpha \,\left[ 
 r_g\sin\delta_g + r_u \sin\delta_u
- i (r_g \cos \delta_g - r_u \cos \delta_u)\right] \;.
\ee
The real and imaginary parts of this relation are each observable,
as they correspond to distinct $f_i$-dependent terms in 
Eq.~(\ref{sumAbarA}). The combination $T_g T_{u}^{\ast}$ can 
be complex, though we assume it to be real for crispness of
discussion. In the imaginary part, we see that 
direct CP violation can exist if the strong phases of $a_j$ vanish,
i.e., if $\delta_u=\delta_g=0$;
merely the difference of $r_g$ and $r_u$ must be non-zero to realize direct
CP violation were $\sin\alpha\ne 0$. If $\delta_j=0$
the strong phase is provided by the resonance width, 
${\rm Im}(f_i f_j^*) \ne 0$.
 Theoretical estimates 
suggest that $r_g$ and $r_u$ are both non-zero and unequal~\cite{ali}. 
In constrast, a partial rate asymmetry can be written as 
\be
|a_g|^2 - |\bar a_g|^2 
= -4 |T_g|^2 r_g \sin\delta_g\,\sin\alpha \;, 
\ee
yielding the familiar result that 
both $r_g$ and $\delta_g$ must be non-zero to yield direct
CP violation were $\sin\alpha\ne 0$. Such conditions are realized in the
real part of Eq.~(\ref{breakdown}) as well, so that the direct
CP-violating observables we propose can be manifest irrespective of 
the strong phases of $a_j$, as they can be non-zero 
were $\delta_j$ either zero or 90 degrees. 
This greater flexibility arises 
as the combination 
$P_g/T_g - P_{u}^{\ast}/T_{u}^{\ast}$ appears in 
Eq.~(\ref{breakdown}), whereas $P_g/T_g - P_{g}^{\ast}/T_{g}^{\ast}$, e.g.,
appears in the partial rate asymmetry. 

Interestingly, 
similar considerations arise in the angular  
analysis of $B\to V_1 V_2$ decays: there, too, 
a CP-odd interference term can beget direct CP violation
in untagged decays~\cite{sinha,snyder}. There are three 
helicity amplitudes, labelled by the helicity $\lambda\in (0,\pm 1)$ 
of either vector meson in $B\to V_1 V_2$ decay. 
Working in a transversity basis~\cite{dunietz}, we can define
the amplitudes $A_\parallel\equiv (A_{+1} + A_{-1})/\sqrt{2}$ and 
$A_\perp\equiv (A_{+1} - A_{-1})/\sqrt{2}$~\cite{cleovv}. 
The full angular distribution of the summed amplitudes for 
$B^0$ and $\bar B^0$ decay permits the extraction of the
imaginary part of the amplitude combinations of Eq.~(\ref{direct}), 
under the identification $a_g \to A_\parallel$, 
$a_u\to A_\perp$, and $a_n \to A_0$. 
Moreover, these untagged 
contributions are insensitive to the strong phase~\cite{valencia}.

The conditions which permit the realization of 
direct CP violation in untagged modes
are quite general. We need only consider self-conjugate final states 
whose resonances encode enantiomeric pair correlations. 
Self-conjugate final states can be realized not only through the 
$b\to d q\bar q$ decays of $B_d$ mesons but also through
the $b\to s q\bar q$ decays of $B_s$ mesons, where 
$q\in u,d,s,c$ quarks. 
The KM picture of CP violation suggests
that direct CP-violating effects ought be suppressed by 
a factor of ${\cal O}(\lambda^2)\sim 1/20$ in $B_s$ meson decay to charmed,
self-conjugate states. Thus
the goals of direct CP violation searches in $B_d$ and $B_s$
meson decays can be distinct. The appearance of direct CP violation
in $B_d$-meson decays would substantiate the KM picture of CP violation,
whereas its appearance in any significant measure in $B_s$ decays 
to charmed final states would signal the presence of
new physics. 
Physics with $B_s$ mesons is important to the future 
B-physics programs at the Tevatron~\cite{Anikeev:2001rk} and at 
the LHC~\cite{Ball:2000ba}. 
The effective tagging efficiency $\epsilon_{\rm eff}$ 
is significantly smaller in a hadronic
environment, 
cf. $\epsilon_{\rm eff} \sim 7\%$~\cite{cdf} with 
$\epsilon_{\rm eff} \sim 27\%$~\cite{Aubert:2002rg,Abe:2002wn} 
at the B-factories, 
so that the untagged studies we propose 
significantly enable direct CP violation searches at these facilities
\footnote{Recall that $\epsilon_{\rm eff}$, 
a conflation of the tagging efficiency
$\epsilon$ and the mistag fraction $w$ given by 
$\epsilon_{\rm eff}=\epsilon (1 -2 w)^2$, drives
the statistical error in an asymmetry measurement 
as per $1/\sqrt{\epsilon_{\rm eff} N}$, where $N$ is the number of
untagged events.}.

Let us enumerate three-, four-, and five-particle final states 
in $B_d$ decay which could yield direct CP violation
in the KM picture. We thus focus on 
$b\to d u \bar u$ and $b\to d c \bar c$ decays, and some possibilities
are given in Table \ref{tab:modes} --- we do not attempt to be exhaustive. 
The CP-enantiomers are useful in the sense we have illustrated
in $B\to \rho\pi$ decay: they permit the formation of 
manifestly CP-odd amplitude combinations 
which can be probed through asymmetries in the population of 
events in the 
regions where the resonances of the CP-enantiomeric pair occur.
We expect the CP-violating effects to be 
larger for broad resonances such as the $\rho$ and $K^*(892)$. 
Note that the final states $K^+ K^- \pi^0$  and 
$K^+ K^- \pi^+\pi^-$, with the CP enantiomers indicated, also lend
themselves to direct CP violation searches in $B_s$ decay. 
Multiparticle final states can support more than one CP-enantiomeric
pair, as illustrated in $B_d\to \pi^+\pi^-\pi^+\pi^-\pi^0$ decay. 
In the case of CP enantiomers which have more than one spin one particle,
as in $(a_1(1260)^+ \rho^- \,,\, a_1(1260)^- \rho^+)$, or which
are not realized by a quasi-two-body decay, 
as in $(\rho^+ \pi^- \pi^+ \pi^-, \rho^- \pi^+ \pi^+ \pi^-)$, 
a caution is in order. For example, 
the presence of two spin-one particles in the final state 
implies that partial waves with $L=0,1$, or $2$ can occur;
the factor $(-1)^L$ impacts the CP of the state. The sum and
difference of the amplitudes associated with 
$B^0 \to a_1(1260)^+ \rho^-$ 
and $\bar B^0 \to a_1(1260)^- \rho^+$ decay 
still yield combinations with definite CP properties 
for any particular $L$, but for $L=0$ or $2$ 
the sum of amplitudes, with a suitable choice of phase 
conventions, does not change sign under CP, whereas for $L=1$
the sum of amplitudes do change sign under CP. 
In either event, for fixed $L$, the CP-odd 
amplitude combination of Eq.~(\ref{breakdown}) appears and drives a
population asymmetry under the exchange of the momentum of a 
$\pi^+$ emerging from the $a_1(1260)^+$ and that 
of the $\pi^-$ from the $\rho^-$ in the region of the Dalitz plot where the 
resonances of the CP-enantiomeric pair occur. States of fixed 
$L$ can be realized through a helicity analysis; the formation
of the $A_\perp$ amplitude, e.g., selects the $L=1$ state~\cite{dunietz}. 
In the absence of a helicity analysis, both 
CP-even and CP-odd contributions are subsumed in ``$g \times u$'' term of
Eq.~(\ref{breakdown}),  so that a population asymmetry 
in this case can exist without direct CP violation. Thus for pairs with
two spin one particles, a helicity analysis is required; similar
considerations apply to pairs for which the decays are not
quasi-two-body in nature --- an ancillary angular analysis is necessary.

\renewcommand{\arraystretch}{1.2}
\begin{table}[hbt]
\begin{center}
\centerline{\parbox{15cm}
{\caption{$B_d$ decays to certain three-, four-, and five-particle, 
self-conjugate final-states and some of the CP-enantiomers they
contain. 
                   \label{tab:modes}}}}
\vspace{.3cm}
\begin{tabular}{||c||c||}
    \hline
    \hline
3-particles  & CP-enantiomers \\
\hline
$\pi^+\pi^-\pi^0$ &  $(\rho^+\pi^- \,,\, \rho^-\pi^+)$ \\
$K^+K^-\pi^0$  & $(K^*(892)^+ K^- \,,\, K^*(892)^- K^+)$ \\
$D^+ D^-\pi^0$ & $(D^*(2010)^+ D^-\,,\, D^*(2010)^- D^+)$ \\
$D^0\bar D^0\pi^0$ & $(D^*(2007)^0 \bar D^0\,,\, \bar D^*(2007)^0 D^0)$ \\
\hline
4-particles  & CP-enantiomers \\
\hline
$\pi^+\pi^-\pi^0 \pi^0$ & $(\rho^+\pi^- \pi^0 \,,\, \rho^-\pi^+ \pi^0)$ 
\footnotemark[1] \\
$\pi^+\pi^-\pi^+ \pi^-$ & $(a_1(1260)^+ \pi^- \,,\, a_1(1260)^- \pi^+)$ \\
$K^+K^-\pi^+ \pi^-$ & 
$(K^*(892)^0 K^-\pi^+ \,,\, {\bar K}^*(892)^0 K^+\pi^-)$ \footnotemark[1] \\
$D^0 \bar D^0 \pi^+ \pi^-$ & $(D^*(2010)^+ \bar D^0 \pi^- \,,\, 
D^*(2010)^- D^0 \pi^+)$\footnotemark[1] \\
\hline
5-particles  & CP-enantiomers \\
\hline
$\pi^+\pi^-\pi^+\pi^- \pi^0$ & $(\rho^+ \pi^-\pi^+\pi^-\,,\, 
\rho^- \pi^-\pi^+\pi^+)$\footnotemark[1] \\
& $(a_1(1260)^+ \pi^- \pi^0\,,\, 
a_1(1260)^- \pi^+\pi^0)$ \footnotemark[1] \\
& $(a_1(1260)^+ \rho^- \,,\, 
a_1(1260)^- \rho^+)$\footnotemark[1] \\
& $(a_0(980)^+ \pi^-\,,\, a_0(980)^- \pi^+)$ \\
& $(b_1(1235)^+ \pi^-\,,\, b_1(1235)^- \pi^+)$ \\
\hline                             
\hline
\end{tabular}
\footnotetext[1]{A helicity and/or angular analysis is required; see text.}
\end{center}
\end{table}

The observation of direct CP violation in B-meson decays in itself
is crucial to establishing 
the mechanism of CP violation.
Nevertheless, we would also like to interpret such results 
in terms of the parameters
of the CKM matrix. An assumption of isospin
symmetry can codify and 
potentially determine the hadronic 
parameters needed to interpret 
the mixing-induced CP-asymmetry 
in $b\to d q \bar q$ transitions to 
charmless final states. 
Relevant to the modes we discuss 
are the isospin-based analyses 
which yield $\sin(2\alpha)$ in 
$B\to\rho\pi$~\cite{LNQS,QS1,QS} and 
$B\to a_1\pi$~\cite{babarbook} decays. 
These analyses, however, do not determine the parameters
necessary to interpret direct CP violation; the terms containing
$\sin\alpha$ and $\cos\alpha$ are multiplied by 
unknown hadronic parameters. Nevertheless, 
were $\sin(2\alpha)$ determined and direct CP violation observed,
the SM value of $\sin\alpha$ 
could be inferred, modulo discrete ambiguities. 
Interpreting direct CP-violating observables
directly in terms of the underlying weak parameters
may not prove possible.  Theoretical
progress has been made, however, in the computation of partial-rate
asymetries in some two-body decays, see, e.g., Refs.~\cite{bbns,hnli}.
Alternatively, more phenomenological treatments indicate that 
the presence of resonances in certain channels can enhance 
the associated partial rate asymmetry~\cite{atwood,ET}
and aid in the extraction of weak phase 
information~\cite{svghocawt}.

We have discussed the conditions under which the rich resonance
structure of hadronic $B$ decays can be exploited to search
for direct CP violation in untagged decays.
Our method is sufficiently general to enable direct CP violation
searches in $B_s$ and $D$ meson decays as well. 
In some channels the untagged search we propose complements 
tagged, time-dependent analyses in $B\to\rho\pi$ and 
$B\to a_1\pi$ decays. Nevertheless, the 
gain in statistical power realized in untagged versus tagged searches,
i.e., roughly a factor of $2$ at the B-factories and of $4$ in a hadronic
environment such as at CDF, argues for a more comprehensive program. 

\medskip
\noindent{\bf Acknowledgements}
S.G. thanks H.R. Quinn for key
discussions and input, L. Dixon and J. Tandean for useful comments, 
and J.D. Bjorken for a helpful conversation.
S.G. acknowledges the SLAC Theory Group for gracious 
hospitality and is supported by
the U.S. Department of Energy under contracts 
DE-FG02-96ER40989 and DE-AC03-76SF00515.


\end{document}